# KAMIWAAI – INTERACTIVE 3D SKETCHING WITH JAVA BASED ON *Cl(4,1)* CONFORMAL MODEL OF EUCLIDEAN SPACE

Eckhard M. S. Hitzer
*Dept. of Mech. Engineering, Fukui Univ.*
*Bunkyo 3-9-, 910-8507 Fukui, Japan.*
*Email: hitzer@mech.fukui-u.ac.jp, homepage: http://sinai.mech.fukui-u.ac.jp/*

**Abstract.** This paper introduces the *new* interactive Java sketching software *KamiWaAi*, recently developed at the University of Fukui. Its graphical user interface enables the user without any knowledge of both mathematics or computer science, to do full three dimensional "drawings" on the screen. The resulting constructions can be reshaped interactively by dragging its points over the screen. The programming approach is new. KamiWaAi implements geometric objects like points, lines, circles, spheres, etc. directly as software objects (Java classes) of the same name. These software objects are geometric entities mathematically defined and manipulated in a conformal geometric algebra, combining the five dimensions of origin, three space and infinity. Simple *geometric products* in this algebra represent geometric unions, intersections, arbitrary rotations *and* translations, projections, distance, etc. To ease the coordinate free and matrix free implementation of this fundamental geometric product, a *new* algebraic three level approach is presented. Finally details about the Java classes of the *new* GeometricAlgebra software package and their associated methods are given. KamiWaAi is available for free internet *download*.

**Key Words:** Geometric Algebra, Conformal Geometric Algebra, Geometric Calculus Software, GeometricAlgebra Java Package, Interactive 3D Software, Geometric Objects

## 1. Introduction

The name "KamiWaAi" of this new software is the Romanized form of the expression in verse sixteen of chapter four, as found in the Japanese translation of the first Letter of the Apostle John, which is part of the New Testament, i.e. the Bible. It simply means "God *is* love." (Please compare the quotation at the end of this introduction.)

The ready-to-use application is written in the platform independent programming language Java. Most computers nowadays already have a Java Runtime Environment (JRE) of the Java 2 Platform, Standard Edition (J2SE) installed. If not, JREs are available for free downloading from the Sun Microsystem homepage[1].

KamiWaAi version 0.0.1 beta is available for free noncommercial use as a zip compressed binary code set of files for download from: http://sinai.mech.fukui-u.ac.jp/gcj/software/KamiWaAi/index.html

This site also features a screenshot (Fig. 1), showing how the KamiWaAi graphical user interface can be used for interactive sketching.

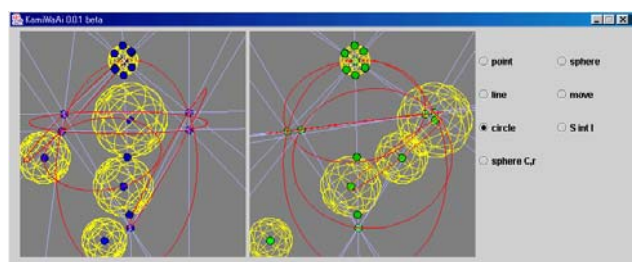

Fig. 1 KamiWaAi screenshot: front (left) and right-side view (center) panels, function radio buttons (right).

The site further gives online access to the KamiWaAi Readme file with information about the naming, warranty, contact for user comments, installation information, brief information on how to use it and references. In the future new versions and detailed documentation will be available from this site.

KamiWaAi has been tested on Windows 98 platforms with the JREs of J2SE version 1.2.2_014 and 1.4.1_01. It should therefore operate across all operating systems (Windows, McIntosh, Unix, Linux, …) without problem and with all JREs currently maintained by Sun Microsystems.

A small photo in the lower right corner of the KamiWaAi homepage shows models of all regular Platonic polyhedral solids: tetrahedron, cube, octahedron, dodecahedron and icosahedron.

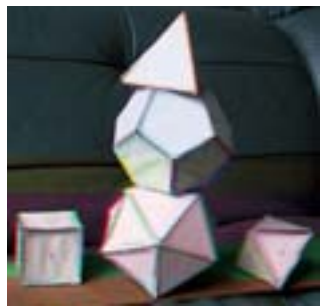

Fig. 2 Platonic polyhedrons.

The reflection groups of these Platonic solids contain the rotational symmetry groups of these solids as subgroups, because the combination of two reflections always yields a rotation.[2],[3] Both the reflection and the rotation symmetry groups of the Platonic solids are of enormous importance for the study of molecular, crystal and lattice symmetries in many fields of science.

The author has already created[4] a series of online applets[5] for the study of the two dimensional symmetry groups of regular polygons (n=1, … ,6). This two dimensional point group applet series features both reflection and rotation symmetries. These applets were designed with the interactive (two dimensional) geometry software Cinderella.[6] The design aims to show how the geometric algebra of vectors in two dimensions represents every of these two dimensional groups just using two physical polygon vectors. It is clear that the two dimensional point symmetry groups are nothing but planar subgroups of full three dimensional crystal groups. Geometric algebra treats these groups using three particular lattice vectors to geometrically represent all reflections and rotations.

The challenge is to visualize this interactively, in order to supply a hands on experience of the great beauty and simplicity of representing the 32 point groups that leave some three dimensional lattice invariant and the full set of 230 space groups, i.e. the complete symmetry groups of three dimensional crystals.

To describe the space groups with full elegance it proves convenient to choose a five dimensional conformal model of the Euclidean three space. Then rotations and translations both become simple monomial (one term) expressions. Similar to projective geometry, which adds one dimension, we add here two dimensions, representing the *origin* and spatial *infinity*. Physicists are by now very familiar with the four dimensional space-time description of special relativity and the fact that null-vectors describe the propagation of light. Here we choose a space with two linearly independent null-directions added to the usual Euclidean space. For a physicist this means simply a signature (+4,-1) Minkowski space.

Then not only the description of crystal symmetries benefits, but we indeed arrive at a new consistent algebra whose elements are in an obvious one-to-one conformal correspondence with real geometric objects like points, lines (and circles), planes (and spheres), volumes, etc. This obviously demands a modern object oriented computer program implementation. With such a fundamental set of encoded geometric objects and their geometrical methods of union, intersection, rotation, translation, etc. the programmer can work free of the restraints of conventional matrix algebra implementations, where coordinates and matrix coefficients often completely hide geometric concepts and invariants.

To show, that this is not pure fantasy, but can be implemented as a matter of principle is the main aim of this paper and of the KamiWaAi software. The backbone of KamiWaAi is therefore a package of repeatedly used GeometricAlgebra objects

(Java classes). These classes are points, lines (circles), planes (spheres) and more general multivectors. It proved practical to hierarchally implement multivectors based on simpler objects of complex numbers (scalars and 5D pseudoscalars), and (complex) quaterions.[7],[8] That way the 32 dimensional multivectors get a natural easy-to-use sub-structure.

It should be obvious that the GeometricAlgebra package maybe used in many other contexts. Wherever the far reaching mathematical apparatus of this algebra is wanted for object oriented computer implementation. This includes diverse areas[2],[3],[13],[17],[18] as simulation, aviation, structural mechanics, navigation, robotics, computer graphics, molecular modeling, solid state physics, cosmology, electrodynamics, gravitation, quantum mechanics, …

From a historical viewpoint five people pioneered this approach (alongside many others): Gottfried W. Leibnitz[9], Sir Rowan Hamilton[7],[10] of Ireland, Herrmann Grassmann[21] of Germany, William K. Clifford[20] of England and the American David Hestenes.[10]

Currently a research group around L. Dorst (Amsterdam) is pursuing a professional Java implementation of geometric algebra. The GeometricAlgebra package of KamiWaAi offers worldwide one of the first implementations of conformal geometric algebra. And to my knowledge KamiWaAi represents worldwide the first exclusively interactive Java software based on conformal geometric algebra. It therefore takes its place at the frontier of exciting new developments in applied mathematics and computer science.

Yet I must caution, that at the moment KamiWaAi is far from optimal concerning aspects of speed optimization. But I regard this as a technical issue and not at all as a principal problem. It will be mended in due time. Another aspect is, that this is one of the first *beta* versions. This means that only very, very few functions are implemented in the 0.0.1 beta version of KamiWaAi and many more functions are scheduled to be implemented soon: E.g. planes, the intersections of spheres, circles, etc, next reflections and rotations, arbitrary choice of angular perspective, etc. Conformal geometric algebra provides straightforward algebraic expressions for all these functions, they just need to be added as methods of the corresponding geometrical objects.

*Advice for users:* If you are primarily interested in knowing how to use KamiWaAi for 3D sketching, just read section 2. Section 3 gives mathematical background information, some of it original. Section 4 is of interest to *programmers*, who want to apply the GeometricAlgebra package for developing other applications or applets.

The name of "KamiWaAi" originates from the Japanese version[11] of: *And we ourselves know and believe the love which God has for us. God is love, and those who live in love live in union with God and God lives in union with them.*[12]

**2. What KamiWaAi Can Do Now**

After starting the application with the command

java –cp . KamiWaAi

from the directory where the object KamiWaAi.class and the package GeometricAlgebra reside, e.g. from c:¥KWA, a new window (compare Fig. 1) will open on the computer screen. The window is composed of three main panels. The left panel shows the front view, the center panel shows he right side view and to the right the available functions are grouped as interactive radio buttons with their labels. In the default configuration the function "point" is activated. I will now quickly step you through the presently available seven functions: point, line, circle, sphere C,r, sphere, move and S int l.

**2.1 Interactively Defining 3D Points with "point"**

If the radio button "point" is activated, the user can define three dimensional points by first clicking on the left panel and next on the center panel. Clicking on the left panel immediately creates a small blue disk at the position of the click. The left panel stays then deactivated (i.e. unreceptive for mouse clicks) until a corresponding right side view point is defined by clicking on the center panel. On the center panel a green disk will appear at the position of the click, exactly in the same height as the blue point just defined on the left panel. Then the center panel becomes deactivated, the left panel becomes activated and a new point can be defined.

Later the underlying representation of these points (vectors) in conformal geometric algebra and by its software implementation in the GeometricAlgebra Java package will be explained later.

## 2.2 Drawing Straight Lines with "lines"

After defining at least two points with the function "points" the user can activate the radio button labeled "line" for drawing straight lines through the points. The choice of points is performed by simply clicking on the small colored disks representing the points in the front and side view panels (left and center panel respectively). On the command line of the shell or MSDOS prompt window the messages "point 1 chosen!" and "point 2 chosen!" will appear successively. As soon as the second message appears a sky blue straight line will be drawn through both points, visible both in the front and in the side view panels. Now new points can be chosen to draw more lines.

It does not matter from which panel the points are chosen, i.e. whether they are chosen both from the same panel view or e.g. first from the left panel and then from the center panel and vice versa.

## 2.3 Drawing Circles with "circle"

Drawing circles after activating the radio button "circle" is very much similar to drawing lines. The only difference is that the user needs to choose a total of three points instead of two. Therefore before being able to create a circle, the user must at least define three points with the function point. The messages "point 1 chosen!", "point 2 chosen!" and "point 3 chosen!" will appear successively on the shell or MSDOS prompt window. After the third message appears, a red circle will be drawn in both views intersecting all three chosen points. Now new points can be chosen to draw more circles. It again does not matter how many of the three points are chosen from the front and/or the side view panels.

## 2.4 Defining a Four Base Point Sphere with "sphere"

First the radio button labeled "sphere" needs to be activated. Then four points need to be chosen similar to clicking the points when drawing lines or circles. Therefore before being able to create a sphere by four base points, the user must at least define four points with the function point. The four messages "point 1 chosen!" …"point 4 chosen!" will appear successively on the shell or MSDOS prompt window. After the fourth one appears, a yellow circle will be drawn, visible both in the front and in the side view panels. Again it does not matter how many points are chosen from the front and/or the side view panel, respectively. Spheres are drawn by their longitudes and latitudes. The longitudinal meridians intersect in two *poles*. The angular position of the poles on a sphere is only for the convenience of displaying them and for later easing the interactive selection of spheres (compare the section on the intersection of spheres and lines.)

## 2.5 Defining a Centered Radius r Sphere with "sphere C,r"

Selecting the radio button "sphere C,r" provides a more convenient way to draw spheres. The user just needs to interactively choose the center point. Immediately after that a small new automatic interactive dialogue window will appear. Entering the desired radius by pressing the numbers on the computer keyboard followed by pressing the ENTER key will cause KamiWaAi to immediately draw a yellow sphere of the desired radius and around the selected center point. The appearance of the sphere is otherwise similar (composed of longitudes and latitudes and with visible *poles*).

As with all other functions, the user can proceed to define more centered radius r spheres with centers and radii at will. Or he can proceed to any of the other implemented functions.

## 2.6 Moving Points and Objects that Depend on them with "move"

Up till now there is no great difference to drawing things on a peace of paper, even if one has to admit that the computer can draw much faster and more precise than sketching by hand usually allows. But the interactive "move" function is so to say the first function that really capitalizes on the ability of a computer to memorize graphical objects and adapt them to changes of the objects which served to define them in the first place. At the moment the *defining objects* are basically the points created with the "point" function.

The first step is to activate the radio button labeled "move". Then the user can click on any point in any of the two panels (front or right side view), keep the mouse button firmly pressed and slowly move it across the screen. The point will follow this motion. After releasing the point again, KamiWaAi will reconstruct all previously present objects, adapting positions, orientations and sizes where it is required. This is certainly something, which needs much more work, fresh sheets of paper and a good understanding of the changing projected appearances if sketched by hand.

The user can use this "move" function visually to interactively explore the dependence of his constructions on the positions of the defining objects (here points) in three dimensions. All objects will follow this motion. (The only exception is mentioned in the next paragraph, but the next program version will eliminate this exception.)

**2.7 Intersecting Spheres with Lines Using "S int l"**

This function needs some explanation. After activating the radio button "S int l" the user is expected to click within three pixels of the line, which he wants to use for the intersection operation. It does not matter from which panel the line is chosen. If the user clicks very close to two lines, KamiWaAi will automatically choose the line whose projection in the panel view is the closest to the clicked position of the panel. The click position has in no way to be a "point" as defined in section 2.1. Any position close enough to the projected view of the straight line will do! If the clicked position is not close enough to any line projection, the computer will wait for another click in order to properly select a line and will give out the information message "Line No. −1 selected." in the shell or MSDOS prompt window. After successfully selecting a line the computer will print the number of the straight line immediately followed by the message "And now select a sphere!"

To select a sphere the user simply has to click on one of the two previously mentioned *poles* of the desired sphere. Remember that a *pole* is marked by the intersection of the longitudinal meridians, which are used to display the sphere. There is a small sensitive area around each *pole* of each sphere in both of the display panels (front and right side). If the clicked position is not close enough to any pole the message "Sphere No. −1 selected." will appear. In this case the user can just continue to click in the vicinity of any pole until the message with the selected sphere number (a positive integer number including zero) will appear in the shell or MSDOS prompt window. Immediately after successfully selecting a sphere the computer will do three actions:

First it will inform the user how many points of intersection of the chosen line and sphere exist: "Two points of intersection!" or "One point of intersection!" or "No intersection!"

In the case of "No intersection!", the computer will immediately restart all over again with the message: "Select a new line!"

Second in the cases of "Two points of intersection!" or "One point of intersection!", KamiWaAi will draw the point(s) of intersection with dark blue disk(s) in the (left) front view panel and with dark green disk(s) in the (central) right side view panel.

Thirdly the computer will generate the shell or MSDOS prompt message: "Select a new line!" KamiWaAi thereby indicates that it is ready for the interactive selection of the next line for the next round of intersection of a sphere and a line.

Here a remark about the *exception* mentioned for "move" operations at the end of the last section is in order. The "S int l" generated intersection points can be used like all other points to define new geometric objects (e.g. lines, circles and spheres). But in the "0.0.1 beta" release of KamiWaAi they will not (yet) follow the move operations of points which cause the defining sphere or line of an intersection point to change its position, orientation or size. *This is an exception that will be remedied in the next release of KamiWaAi!*

Now all presently implemented seven functions are explained in detail and the interested reader is hereby invited to start exploring three dimensional geometry on his own using KamiWaAi.

Another note of caution is in order: The display is not permanent. That is, if the KamiWaAi display window is overshadowed by any other window, the overshadowed portion will be lost until another move operation takes place. A temporary *workaround* is to just define a new point and move it with "move" as described in section 2.6. This *temporary loss of display* will also be dealt with in the next release. As a reason I submit, that KamiWaAi is in fact the first ever Java program which I write from scratch. I naturally welcome advise on such technical issues.

In the next two sections first the background in the theory of conformal geometrical algebra will be outlined. Then the Java package GeometricAlgebra, which implements this algebra in the form of new object oriented Java classes and their methods will be documented. This description inform about the major classes, list their most important methods, specifying their input and output. The reason for first dealing with the conformal geometrical algebra background is, that with this knowledge at hand, the way the classes are defined and the methods they posses will become obvious and natural.

## 3. The Conformal Geometric Algebra of Origin, Euclidean Three Space, and Infinity

In this section I will give a straightforward description of essential elements and relations of the five dimensional conformal geometric algebra concerned. For further details and some proves, I refer the interested readership to publications.[13]-[15] However it appears to me that the essential employment of a three level sub-algebra structure is a new and original approach to the subject.

### 3.1 Multivectors and General Multivector Product

Fine introductory descriptions of the geometric algebra of the Euclidean (two and) three space can be exist.[15],[16] This algebra consisting of grade 0 scalars α, three linearly independent orthonormal grade 1 vectors $\{\vec{e}_1, \vec{e}_2, \vec{e}_3\}$, three linearly independent grade 2 bivectors

$$\{i_1, i_2, i_3 \equiv \vec{e}_2\vec{e}_3,\ \vec{e}_3\vec{e}_1,\ \vec{e}_1\vec{e}_2\} \tag{1}$$

(equivalent to the *i,j,k* of Hamilton's quaternions and understood as the oriented side areas of a unit cube) and one square minus one pseudoscalar three-volume *i* with square minus one $i^2 = -1$. This $2^3 = 8$ dimensional geometric algebra has therefore the algebraic basis:

$$\{1,\ \vec{e}_1, \vec{e}_2, \vec{e}_3,\ i_1, i_2, i_3,\ i = \vec{e}_1\vec{e}_2\vec{e}_3,\}. \tag{2}$$

This geometric algebra of the Euclidean three space in itself is already of great interest for mechanics, robotics, etc.[10],[16]-[18] But without pause, I will now add the two linearly independent null-vectors for the *origin* $\{\vec{\bar{n}}\}$ and for *infinity* $\{\vec{n}\}$ to give the full set of five basic vectors:

$$\{\vec{\bar{n}},\ \vec{e}_1, \vec{e}_2, \vec{e}_3,\ \vec{n}\}. \tag{3}$$

Equipped with the *geometric product* invented by H. Grassmann[19] and W.K. Clifford[20] and augmented by product rules for higher grade objects we get the full $2^5 = 32$ dimensional conformal geometric algebra with its general multivector product. The bilinear and associative but not commutative *geometric product* was first introduced by Grassmann to integrate Hamilton's quaternions[10] into his own extensive algebra. Independent of the dimensions of the space concerned, it unifies the conventional scalar product of vectors and Grassmann's earlier antisymmetric outer bivector product of vectors.

$$\vec{a}\vec{b} = \vec{a} \cdot \vec{b} + \vec{a} \wedge \vec{b}. \tag{4}$$

Before tabulating the full multiplication table of all 32 basic elements it proves to be very instructive and immediately useful to first look at the product table of the origin and infinity {null-vectors $\vec{\bar{n}}$ and $\vec{n}$}. A physicist can easily derive these relationships by adding two vectors of opposite signature ($\vec{e}_0^{\ 2} = -1, \vec{e}_4^{\ 2} = +1$) to the basis of the Euclidean three dimensional space and defining the two null-vectors in terms of these two extra vectors:

$$\vec{\bar{n}} \equiv \vec{e}_0 + \vec{e}_4,\ \vec{n} \equiv \tfrac{1}{2}(\vec{e}_0 - \vec{e}_4). \tag{5}$$

We then have the following fundamental null-vector products:

$$\vec{n}^2 = \vec{n}\vec{n} = 0, \vec{\bar{n}}^2 = 0, \tag{6}$$

$$\vec{n}\vec{\bar{n}} = -1 + \vec{n} \wedge \vec{\bar{n}}, \vec{\bar{n}}\vec{n} = -1 - \vec{n} \wedge \vec{\bar{n}}, \tag{7}$$

$$\vec{n} \cdot \vec{\bar{n}} = \vec{\bar{n}} \cdot \vec{n} = -1. \tag{8}$$

Defining $N \equiv \vec{n} \wedge \vec{\bar{n}}$ we further obtain

$$N\vec{\bar{n}} = -\vec{\bar{n}}N = \vec{\bar{n}}, N\vec{n} = -\vec{n}N = -\vec{n}, \tag{9}$$

$$N^2 = NN = 1. \tag{10}$$

All these products can be conveniently summarized in the following sub-algebra $\{1, \vec{n}, \vec{\bar{n}}, N\}$ multiplication table:

Table 1 Sub-algebra $\{1, \vec{n}, \vec{\bar{n}}, N\}$ multiplication table.

|  | 1 | $\vec{n}$ | $\vec{\bar{n}}$ | $N$ |
|---|---|---|---|---|
| 1 | 1 | $\vec{n}$ | $\vec{\bar{n}}$ | $N$ |
| $\vec{n}$ | $\vec{n}$ | 0 | $-1+N$ | $\vec{n}$ |
| $\vec{\bar{n}}$ | $\vec{\bar{n}}$ | $-1-N$ | 0 | $-\vec{\bar{n}}$ |
| $N$ | $N$ | $-\vec{n}$ | $\vec{\bar{n}}$ | 1 |

This table will indeed provide a very convenient top level algebraic sub-structure for the complete multiplication table of the conformal geometric algebra.

Next the quadruple $\{1, i_1, i_2, i_3\}$ has the following quaternionic multiplication table:

Table 2 Quaternionic sub-algebra multiplication table.

|  | 1 | $i_1$ | $i_2$ | $i_3$ |
|---|---|---|---|---|
| 1 | 1 | $i_1$ | $i_2$ | $i_3$ |
| $i_1$ | $i_1$ | -1 | $i_3$ | $-i_2$ |
| $i_2$ | $i_2$ | $-i_3$ | -1 | $i_1$ |
| $i_3$ | $i_3$ | $i_2$ | $-i_1$ | -1 |

I further give an explicit expression for the five dimensional conformal algebra pseudoscalar $I$ as:

$$I = \vec{e}_1\vec{e}_2\vec{e}_3 N = iN, \quad I^2 = II = -1. \tag{11}$$

Now everything is ready to give the complete list of all 32 linearly independent basis elements of the five dimensional conformal *algebra* in Table 3.

Table 3 All 32 linearly independent basis elements of the 5D conformal algebra. Grades in first line.

| 0 | 1 | 2 | 3 | 4 | 5 |
|---|---|---|---|---|---|
| 1 | $\vec{e}_1 = -Ii_1N$ | $i_1$ | $Ii_1 = -\vec{e}_1N$ | $i_1N$ | $I$ |
|  | $\vec{e}_2 = -Ii_2N$ | $i_2$ | $Ii_2 = -\vec{e}_2N$ | $i_2N$ |  |
|  | $\vec{e}_3 = -Ii_3N$ | $i_3$ | $Ii_3 = -\vec{e}_3N$ | $i_3N$ |  |
|  | $\vec{n}$ | $\vec{e}_1\vec{n} = Ii_1\vec{n}$ | $i_1\vec{n}$ | $I\vec{n} = -\vec{e}_1\vec{e}_2\vec{e}_3\vec{n}$ |  |
|  |  | $\vec{e}_2\vec{n} = Ii_2\vec{n}$ | $i_2\vec{n}$ |  |  |
|  |  | $\vec{e}_3\vec{n} = Ii_3\vec{n}$ | $i_3\vec{n}$ |  |  |
|  | $\vec{\bar{n}}$ | $\vec{e}_1\vec{\bar{n}} = -Ii_1\vec{\bar{n}}$ | $i_1\vec{\bar{n}}$ | $I\vec{\bar{n}} = \vec{e}_1\vec{e}_2\vec{e}_3\vec{\bar{n}}$ |  |
|  |  | $\vec{e}_2\vec{\bar{n}} = -Ii_2\vec{\bar{n}}$ | $i_2\vec{\bar{n}}$ |  |  |
|  |  | $\vec{e}_3\vec{\bar{n}} = -Ii_3\vec{\bar{n}}$ | $i_3\vec{\bar{n}}$ |  |  |
|  |  | $N$ | $i = IN = \vec{e}_1\vec{e}_2\vec{e}_3$ |  |  |

Table 3 features grade by grade one scalar, five vectors, 10 bivectors and the dual 10 trivectors, 5 quadrivectors and the 5D pseudoscalar $I$. The grades are given in the top line of Table 3. The central vertical dividing line between the grades 2 and 3 indicates that all elements on the right are dual to corresponding elements on the left and vice versa. Duality here simply means multiplication with the pseudoscalar $I$. The geometric product is associative, but in general not commutative. Yet scalars 1 and pseudoscalars $I$ commute with all elements and form because of Eq. (11) a sub-algebra isomorphic to complex numbers (Table 4.)

Table 4 $\{1, I\}$ subalgebra isomorphic to complex numbers.

|   | 1 | $I$ |
|---|---|---|
| 1 | 1 | $I$ |
| $I$ | $I$ | -1 |

Please note that our pseudoscalar $I$ has a definite geometric meaning as the product (11) of the three dimensional Euclidean pseudoscalar $i$, as defined in Eq. (2) times the two dimensional origin-infinity null-subspace pseudoscalar $N$.

With the aim of rendering the full multiplication table of all elements listed in Table 3, I grouped the 32 terms into four "complex quaternions" with the help of the subalgebras of Tables 2 and 3.

By complex quaternion I mean the following expression:

$$q = q_s + q_v, \text{ with } q_s = q_{sr} + Iq_{si}, \text{ and}$$

$$q_v = q_1 + q_2 + q_3 = (q_{1r} + Iq_{1i})i_1 + (q_{2r} + Iq_{2i})i_2 + (q_{2r} + Iq_{2i})i_2. \tag{12}$$

The eight coefficients $\{q_{sr}, q_{si}, q_{1r}, q_{1i}, q_{2r}, q_{2i}, q_{3r}, q_{3i}\}$ are all real scalars. Using the multiplication Table 2 the conformal geometric multiplication of two complex quaternions $p$ and $q$ yields:

$$pq = p_s q_s - (p_1 q_1 + p_2 q_2 + p_3 q_3) + (p_3 q_2 - p_2 q_3)i_1 + (p_1 q_3 - p_3 q_1)i_2 + (p_2 q_1 - p_1 q_2)i_3 \tag{13}$$

which is a new quaternion with its "complex" scalar part in the first right hand side line of Eq. (13) and its "complex" bivector part in the second line. The first term $p_s q_s$ of the "complex" scalar part e.g. simply is according to Table 4:

$$p_s q_s = p_{sr} q_{sr} - p_{si} q_{si} + I(p_{sr} q_{si} + p_{si} q_{sr}). \tag{14}$$

A general (up to 32 element) *multivector* element of the conformal geometric algebra, defined by adding scalar multiples of all elements of grades one to five listed in Table 3, can now be rewritten in a very elegant way as:

$$m = q + q_n \vec{n} + q_{\bar{n}} \vec{\bar{n}} + q_N N. \tag{15}$$

$m$ comprises indeed 32 degrees of freedom, because each of the four "complex quaternions" $\{q, q_n, q_{\bar{n}}, q_N\}$ has itself eight real scalar degrees of freedom as shown in Eq. (12). The multiplication of two multivectors $m, m'$ is now simply achieved using Table 1 and the multiplication of the four complex quaternion coefficients according to Eq. (13):

$$mm' = qq' + q_N q'_N - q_n q'_{\bar{n}} - q_{\bar{n}} q'_n + (qq'_n + q_n q' + q_n q'_N - q_N q'_n)\vec{n}$$

$$+ (qq'_{\bar{n}} + q_{\bar{n}} q' - q_{\bar{n}} q'_N + q_N q'_{\bar{n}})\vec{\bar{n}} + (qq'_N + q_N q' + q_n q'_{\bar{n}} - q_{\bar{n}} q'_n)N. \tag{16}$$

It is obvious that this 16 term expression is much easier to handle than a full 32 by 32 = 1024 term expression involving all real 64 degrees of freedom of the two multivectors $m, m'$. It would indeed make no sense to provide such an extensive listing of terms, because we would necessarily be lost in understanding their geometric significance and it would be hard for a programmer to correctly program the 1024 element matrix without error. Opposed to that our three level hierarchy of the full multivector algebra and its subalgebras of complex quaternions and complex numbers also provides a straightforward and absolutely error free way of implementing the full product of Eq. (16), by first implementing the product of complex numbers as in Eq. (14), then implementing the product of quaternions of Eq. (13) and only finally the multivector product of Eq. (16) at the top of this three level hierarchy.

Even the reader void of any knowledge of Java will therefore anticipate that we will have three fundamental objects of complex number (ComplexNumber.class), complex quaternion (ComplexQuat.class) and multivector (MultiVector.class). The complex numbers consist of two dimensional arrays of real scalars double[2], the complex quaternions of four dimensional arrays of complex numbers ComplexNumber[4] and the multivectors of four dimensional arrays of quaternions ComplexQuat[4]. "double" indicates the decimal point real number type variables of Java. Naturally each object has its own method for addition [add()], subtraction [sub()], scalar multiplication [multSc()] and multiplication [mult()], given in Eqs. (14), (13) and (16). In the Java source code we will therefore be able to multiply multivectors m and mp by simply writing:

m.mult(mp);

automatically using the correct multiplication method mult() of the instance m of the MultiVector.class object.

The hardest part of the conformal geometric algebra foundation is now established. Next we need to ask ourselves how to assign meaningful geometric entities (like spatial 3D points, lines, circles, spheres, etc.) to certain multivectors and how to let these geometric entity objects interact with each other. With interactions I mean geometrical unions of (point, line and plane) subspaces, intersections (of lines, circles, planes, spheres, etc.), projections, etc., i.e. geometric set theoretic operations keeping track of the dimensions involved. Other operations of interest will be translations, rotations, dilations and reflections, etc.

It is straightforward to answer the first question for meaningful identification of geometric entities (objects) in the conformal geometric algebra. But in order to establish the geometrical object interactions and operations it proves very convenient to first introduce some grade dependent operations in the multivector algebra and to derive a handful of formidable products from the general multivector product of Eq. (16).

The strategy persued in the following four sections will therefore *first* introduce basic indentifications of geometric entities used by KamiWaAi in its present version. *Second* I will give an overview of important grade dependent changes of multivectors and *third* I will show how these grade dependent operations on multivectors can be used to modify and combine multivectors in derived products. And *fourth* this will enable us to write down very elegant monomial (one term) expressions for the desired geometric object interactions (unions, …) and operations (translations, …).

**3.2 Basic Geometric Entities Identified in the Conformal Geometric Algebra**

**3.2.1 Points**

The first kind of geometric entities we are concerned with are points. We are familiar with the representation of points by position vectors in a three dimensional Euclidean vector space. A three dimensional vector space in our conformal algebra is given by the subspace characterized by the three vectors $\vec{e}_1, \vec{e}_2, \vec{e}_3$ or by their product $i = \vec{e}_1 \vec{e}_2 \vec{e}_3$. Any linear vector combination

$$\vec{x} = x_1 \vec{e}_1 + x_2 \vec{e}_2 + x_3 \vec{e}_3 . \qquad (17)$$

of these three vectors corresponds one to one with a point in a three dimensional Euclidean vector space. The three real scalars $\{x_1, x_2, x_3\}$ are called the coordinates of the point. In order to reap some benefit from our five dimensional conformal origin-space-infinity algebra, we now modify the definition (17) to represent points as conformal vectors $\vec{X}$ with additional origin and infinity components:

$$\vec{X} = \vec{x} + \frac{1}{2} x^2 \vec{n} + \vec{\bar{n}} . \qquad (18)$$

Two remarks are in order: First, if $\vec{x} = 0$ we get $\vec{X} = \vec{\bar{n}}$. This shows why the vector $\vec{\bar{n}}$ is said to represent the origin. If $\vec{x}$ becomes very long, the square $x^2 = \vec{x}\vec{x} = \vec{x} \cdot \vec{x}$ will dwarf the two other terms in (18). This is why the vector $\vec{n}$ is associated with spatial infinity. Second, it is very easy to switch between (17) and (18). To get from (17) to (18) we simply add the two last terms on the right hand side of equation (18). To get back to (17) we simply strip away the two origin and infinity vector parts of (18) to obtain (17) [projection into the Euclidean three space, also called *conformal split* in special relativity.]

The reason for the particular form of (18) can be easily seen by performing the product of two conformal points $\vec{A}, \vec{B}$

$$\vec{A}\vec{B} = \vec{A} \cdot \vec{B} + \vec{A} \wedge \vec{B}. \tag{19}$$

Using Table 1 we get for the scalar inner product part

$$\vec{A} \cdot \vec{B} = -\frac{1}{2}(\vec{a} - \vec{b})^2. \tag{20}$$

This is of great practical interest, because we see that the inner product of conformal vectors (18) simply gives us the distance between the corresponding Euclidean three space vectors $\vec{a}, \vec{b}$. The outer product part of (19) gives

$$\vec{A} \wedge \vec{B} = \vec{a} \wedge \vec{b} + \frac{1}{2}(b^2\vec{a} - a^2\vec{b}) \wedge \vec{n} + (\vec{a} - \vec{b}) \wedge \bar{\vec{n}} + \frac{1}{2}(a^2 - b^2)N \equiv B - \frac{1}{2}\vec{v} \wedge \vec{n} - \vec{u} \wedge \bar{\vec{n}} + \frac{1}{2}\gamma N. \tag{21}$$

$\vec{A} \wedge \vec{B}$ of (21) allows us to recover the original conformal vectors $\vec{A}, \vec{B}$:

$$\sigma \equiv \frac{1}{2}\gamma^2 - \vec{u} \cdot \vec{v}, \quad \rho \equiv \sqrt{\sigma^2 - u^2 v^2}, \quad a^2 = (\sigma + \rho)/u^2,$$

$$b^2 = (\sigma - \rho)/u^2, \quad \vec{a} = (a^2\vec{u} + \vec{v})/\gamma, \quad \vec{b} = (b^2\vec{u} + \vec{v})/\gamma, \tag{22}$$

and by inserting $\vec{a}, a^2, \vec{b}, b^2$ into (18). Equation (22) can be proved by insertion, yet the actual derivation is a direct geometric algebra calculation. We learn therefore that the outer product part $\vec{A} \wedge \vec{B}$ of the geometric product of geometric vectors completely preserves the information about where to find each of the two points in Euclidean three space. Extracting the information of the constituting points out of a bivector like $\vec{A} \wedge \vec{B}$ will have further applications when intersecting e.g. straight lines and spheres, etc.

Let me finally remark that according to Eq. (20) the geometric product of a conformal vectors with itself becomes zero (or null):

$$\vec{A}\vec{A} = \vec{A} \cdot \vec{A} = -\frac{1}{2}(\vec{a} - \vec{a})^2 = 0. \tag{23}$$

This shows that the special definition (18) not only leads to the elegant product form for calculating Euclidean distances of Eq. (20) but also implies that the conformal vectors of form (18) are themselves zero square vectors, i.e. *null-vectors*.

To represent the Euclidean three space by conformal vectors with square zero may at first seem rather artificial. But already the scalar product result of Eq. (20) and the possibility of Eq. (22) to fully extract the original vectors show that we earn a number of important advantages. There are more to come!

### 3.2.2 Circles and Lines

A volume element characterizes the space of which it is part. A line vector $\vec{u}$ in Euclidean three dimensional space gives a way of testing whether any other vector $\vec{x}$ is parallel to the line or not:

$$\vec{u}, \vec{x} \text{ parallel} \iff \vec{u} \wedge \vec{x} = 0. \tag{24}$$

The geometric interpretation of the outer product $\vec{a} \wedge \vec{b}$ of two vectors in three dimensions is the area swept out by displacing one the first vector $\vec{a}$ along the second $\vec{b}$ and vice versa (up to a significant sign change). If we take a third vector $\vec{x}$ and sweep the area of the outer product of the first two along this third, we get a volume, if the third vector is not in the plane of the first two.

(Compare the animated and interactive illustrations of the outer product.[5]) If $\vec{x}$ happens to be in the plane of the first two, we must have for the outer product volume

$$\vec{a} \wedge \vec{b} \wedge \vec{x} = 0 . \tag{25}$$

This can be generalized. The outer product of $d = 1, 2, \ldots$ or 5 linearly independent vectors $\vec{a}_1 \wedge \ldots \wedge \vec{a}_d$ will identify a $d$ dimensional sub-space of the conformal vector space spanned by the basis (3). The test, whether any other vector $\vec{x}$ is part of this $d$ dimensional subspace or not is, whether the outer product with $\vec{x}$ vanishes:

$$\vec{x} \in d\text{-subspace} \Leftrightarrow \vec{a}_1 \wedge \ldots \wedge \vec{a}_d \wedge \vec{x} = 0. \tag{26}$$

After considering the outer product $\vec{A} \wedge \vec{B}$ of two conformal vectors in the last subsection, it is perfectly natural to ask what a conformal three-volume $V \equiv \vec{A}_1 \wedge \vec{A}_2 \wedge \vec{A}_3$ constructed from three conformal points $\vec{A}_1, \vec{A}_2, \vec{A}_3$ corresponds to in the three dimensional Euclidean subspace. I give the answer without prove: It corresponds to a (generalized) circle passing through $\vec{a}_1, \vec{a}_2$ and $\vec{a}_3$. The conformal trivector $V$ encodes all information about this circle: its center point $\vec{C}$, its radius $r$ and its plane bivector $B$.
Then the radius is

$$r^2 \equiv -\frac{VV}{(V \wedge \vec{n})^2} . \tag{27}$$

The plane $B$ is characterized by simply taking the bivector coefficient of $\vec{\vec{n}}$ in $V \equiv B\vec{\vec{n}} + \ldots$ and the conformal center point $\vec{C}$ is given by

$$\vec{C} = -\vec{F}/(\vec{F} \cdot \vec{n}) + \frac{1}{2} r^2 \vec{n} , \tag{28}$$

with $\vec{F} = BNV$. The 3D Euclidean spatial components of $\vec{C}$ give according to Eq. (18) the vector $\vec{c}$.
But what happens if the radius of Eq. (27) becomes infinite, i.e. if

$$V \wedge \vec{n} = 0 \ ? \tag{29}$$

Well a circle with infinite radius has zero curvature and is therefore nothing but a straight line. Comparing the subspace criterion Eq. (26) with Eq. (29) we find that in this case the infinity vector $\vec{n}$ will be part of the *generalized* circle. This is absolutely natural, since we expect that a straight line passes through infinity. This gives us an easy recipe for creating lines: Simply take the outer product of two finite conformal points on the line $\vec{A}_1 \wedge \vec{A}_2$ and take the outer product with the infinity vector $\vec{n}$:

$$V_{line} = \vec{A}_1 \wedge \vec{A}_2 \wedge \vec{n} . \tag{30}$$

It also constitutes an easy test to see if three points are collinear (i.e. all on one straight line) or not:

$$V \wedge \vec{n} \equiv \vec{A}_1 \wedge \vec{A}_2 \wedge \vec{A}_3 \wedge \vec{n} = 0 . \tag{31}$$

**3.2.3 Spheres and Planes**
The transition from circles to spheres happens exactly as intuition may lead to expect. A sphere is characterized by the outer product of four conformal points:

$$V \equiv \vec{A}_1 \wedge \vec{A}_2 \wedge \vec{A}_3 \wedge \vec{A}_4. \tag{32}$$

The radius $r$ is now given by (mark the sign!)

$$r^2 \equiv \frac{VV}{(V \wedge \vec{n})^2} \tag{33}$$

and becomes infinite for $V \wedge \vec{n} = 0$, i.e. again if the infinity vector $\vec{n}$ is "on" the sphere. A sphere with infinite radius (zero curvature) passing through infinity is nothing but a plane. The $V$ of Eq. (32) therefore characterizes both spheres and planes. The test is again performed as in Eq. (29), by performing the outer product with $\vec{n}$. For a genuine sphere with finite radius we can immediately write down the conformal center

$$\vec{C} = -IV/(2r^3) + \frac{1}{2}r^2\vec{n}. \tag{34}$$

At the moment I just state this as a "conjecture", albeit one that has shown to work well, when implemented as Java object method. Again Eq. (18) tells us how to convert the conformal center of Eq. (34) to the Euclidean 3D space center $\vec{c}$ of the sphere.

The formulas of Eqs. (32), (33) and (34) are very simple and very easy to implement with a computer program, which is able to multiply conformal multivectors. I think this already reveals some of the true geometric computation power of the conformal model and again, more is to come, especially in section 3.5.

### 3.3 Grade Sensitive Modifications of Multivectors

In this subsection I will briefly introduce basic grade sensitive modifications of multivectors,[22]-[24] which are highly useful for deriving multivector products of special significance and are frequently applied in the manipulation of multivectors: *grade selection,* the two conjugations of (the antiautomorphic involution) *reversion* and (the automorphic) *grade involution.*

Any conformal multivector can be written in terms of its six grade parts

$$m = \langle m \rangle_0 + \langle m \rangle_1 + \langle m \rangle_2 + \langle m \rangle_3 + \langle m \rangle_4 + \langle m \rangle_5, \tag{35}$$

where the grade indexes of the *grade selectors* $\langle m \rangle_g$ indicate the grades $g$ as tabulated in Table 3. As an example the scalar product of Eq. (20) has grade 0, all vectors in Eqs. (17), (18), ... have grade 1, the outer product of two vectors of Eq. (21) has grade 2, outer products of three vectors of Eq. (30) have grade 3, the outer products of four vectors in Eq. (32) have grade 4 and the five dimensional pseudoscalar of Eq. (11) has grade 5. Grade selection applied to the geometric product [compare Eq. (4)] of vectors results in

$$\vec{a}\vec{b} = \vec{a} \cdot \vec{b} + \vec{a} \wedge \vec{b} = \langle \vec{a}\vec{b} \rangle_0 + \langle \vec{a}\vec{b} \rangle_2, \tag{36}$$

which is another way of noting that the scalar product part is scalar (of grade 0) and that the outer product part is a bivector (of grade 2).

I know immediately apply grade selection in order to define the *reversion* (indicated by the tilde) as

$$\widetilde{m} = \langle m \rangle_0 + \langle m \rangle_1 - \langle m \rangle_2 - \langle m \rangle_3 + \langle m \rangle_4 - \langle m \rangle_5. \tag{37}$$

The name reversion refers to the fact that the reverse order of an outer product of $g$ linear independent vectors is its reversion as defined in Eq. (37). E.g. according to Table 3 we have for $i = \langle i \rangle_3$

$$\widetilde{i} = \vec{e}_3 \vec{e}_2 \vec{e}_1 = \vec{e}_2 \vec{e}_1 \vec{e}_3 = -\vec{e}_1 \vec{e}_2 \vec{e}_3 = -i = -\langle i \rangle_3. \tag{38}$$

Finally the *grade involution* (or main involution) means to change the sign of all odd grade parts, indicated by a small hat

$$\hat{m} = \langle m \rangle_0 - \langle m \rangle_1 + \langle m \rangle_2 - \langle m \rangle_3 + \langle m \rangle_4 - \langle m \rangle_5. \tag{39}$$

The remark is in order that the symbols for grade selection, reverse and grade involution sometimes differ depending on the authors.

Knowing about grade selection, reversion and grade involution we are ready to derive some further useful products from the general multivector product (16).

**3.4 Useful Products Derived from the General Multivector Product**

We already know besides the general multivector product about the scalar product of vectors as in Eq. (20) and about outer products of two or more vectors in Eqs. (21), (30), (32), etc. But it is important to find out whether the general multivector product has some invariant parts, which are of geometric significance and which may for example generalize the notions of scalar product and outer product to multivectors. The answer to this question is an affirmative yes! Researchers in this field have developed a small "zoo" of products derived from the general multivector product.[22],[24],[25]

I attempt here only to name but a few, which are already implemented in the Java package GeometricAlgebra as methods of the object MultiVector.class. These are general scalar and outer products, the socalled left and right contractions and the general scalar magnitude of multivectors. Now the grade sensitive multivector modifications of the preceding section become important. The general multivector product in Eq. (16) is not indicated by any product symbol. It is always assumed if two multivectors are written one after the other (juxtaposition). All derived products have their special symbols, which sometimes differ, depending on the authors.

The *scalar product* of two multivectors $m, m'$ is defined as the scalar (grade zero) part of the general multivector product in Eq. (16):

$$m * m' \equiv \langle mm' \rangle_0 . \tag{40}$$

The Java method ScProd() implementation is therefore utterly simple: The method first applies the general mult() method and than applies the getGrade(0) method to the result.

The *outer product* of two grade selected parts $\langle m \rangle_r, \langle m' \rangle_s$ is given by

$$\langle m \rangle_r \wedge \langle m' \rangle_s \equiv \langle \langle m \rangle_r \langle m' \rangle_s \rangle_{s+r} . \tag{41}$$

This means to first select the grade r part $\langle m \rangle_r$ of $m$, then to select the grade s part $\langle m' \rangle_s$ of $m'$, to compute the full multivector product of $\langle m \rangle_r$ and $\langle m' \rangle_s$ according to Eq. (16) and finally take the grade (r+s) part of the result. This is precisely the way the computer does it, using the methods getGradeI on $m$, and getGrade(s) on $m'$, then the method mult() on the two grade parts and finally the method getGrade(r+s) on the result of the multiplication.

The *general outer product* of two multivectors is then easily defined as the sum over the outer products of all grade parts of the two factors (bilinearity of the outer product):

$$m \wedge m' \equiv \sum_{r=0}^{5} \sum_{s=0}^{5} \langle m \rangle_r \wedge \langle m' \rangle_s . \tag{42}$$

In the past some researchers in this field used a symmetric inner product of multivectors, which is not to be confused with the scalar product of Eq. (40). But this introduced some complications.[25] It therefore seems to be more consistent to continue to use instead two products which are called *left* and *right contractions.*

The *left contraction* is to be understood as contracting a multivector of lower grades from the *left* onto a multivector of higher grades on the right.

$$m \rfloor m' \equiv \sum_{r,s=0}^{5} \langle \langle m \rangle_r \langle m' \rangle_s \rangle_{s-r} . \tag{43}$$

The calculation means to first extract the grade r part of *m* and the grade s part of *m'*, next to use the general multivector product of Eq. (16) to pair wise compute all products of the grade parts and then to select the (*s-r*) parts of the products for the final summation. Trying to left contract a higher grade part from the left side with a lower grade part on the right side of this product (i.e. for *s<r*) will simply produce zero, since in geometric algebra no negative graded parts exist. For the case of two vectors the left contraction is identical to the scalar product of Eq. (40).

The *right contraction* is defined in analogy to the left contraction, just that now the contractions has to be seen as contracting lower grade elements from the *right* side onto higher grades elements on the left side of the product.

$$m \,\lrcorner\, m' \equiv \sum_{r,s=0}^{5} \left\langle \langle m \rangle_r \langle m' \rangle_s \right\rangle_{r-s}. \tag{44}$$

As expected trying to right contract a higher grade element from the right onto a low grade element to the left (i.e. $s>r$) yields zero. For the case of two vectors the right contraction is again equal to the scalar product of Eq. (40). For $r = s$, the right contraction and the left contraction give the same scalar valued results.

There is a natural geometric meaning to left and right contractions. If we think in the terms of projection, it is impossible to project e.g. an area onto a mere line, that is a higher dimensional entity cannot be *contained* in a lower dimensional entity.

### 3.5 Manipulations of Basic Geometric Entities

A common observation is that we can assign certain magnitudes in geometry: lenth, area, volume, … And as we expect therefore, conformal geometric algebra provides a straightforward method of calculation for this. The square of the real scalar magnitude of a multivector $m$ is

$$|m|^2 \equiv \widetilde{m} * m, \tag{45}$$

i.e. the scalar product of Eq. (40) of the reverse (37) of a multivector with $m$ itself. The product in Eq. (45) will be positive for any non-zero $m$ from any positive definite sub-algebra of the conformal geometric algebra. In particular the three dimensional Euclidean subalgebra (2) contains only positive definite elements. Naturally this condition is obviously not fulfilled for the null-vectors of origin, infinity, for the conformal points (18) nor for algebraic elements that contain *one* (but not both) of the origin and infinity null-vectors as a vector factor. Though this may not be our daily experience, special relativists are used to such facts, which are constituent for the very structure of our space-time.

Given that we have a positive definite sub-algebra multivector, we can proceed to normalize it by dividing it by its magnitude

$$m \to \frac{m}{|m|}. \tag{46}$$

This normalized multivector now has according to Eqs. (45) and (46) the magnitude 1 (one).

Apart from allowing us to define the notion of *angle* between multivectors, another use for the magnitude is the definition of an *inverse* element for positive definite single grade multivectors $\langle m \rangle_r$ (sometimes called *blades*) with respect to the general multivector product of Eq. (16):

$$\langle m \rangle_r^{-1} \equiv |\langle m \rangle_r|^{-2} \langle \widetilde{m} \rangle_r. \tag{47}$$

I emphasize that in particular each element listed in the geometrical algebra of Euclidean three space (2) has a multiplicative inverse. We can therefore now *divide* not only by scalars, but also by vectors (!), bivectors and trivectors. It is also possible to divide by $I$ and $N$ of Table 3. The only aspect we need to take care of is that division from the right and division from the left are no longer the same thing, because geometric products are generally not commutative. This *division* property is of immense value for the solution of algebraic equations and it is something that traditional vector algebra completely misses out on. This last fact is quite regrettable, regarding the historical development of the teaching of mathematics and should definitely be remedied as soon as possible.

In general it is possible to calculate integer powers of multivectors by just multiplying them $k$ times:

$$m^k \equiv \underbrace{mm\ldots m}_{k}. \tag{47}$$

Utilizing this we can define the exponential of a multivector m using the familiar power series expansion

$$\exp(m) \equiv \sum_{k=0}^{\infty} \frac{(-1)^k}{k!} m^k. \tag{48}$$

As stated earlier, Hamilton's quaternions are isomorphic to the set (1) adding the scalar 1. This is demonstrated in detail by Table 2. It is therefore absolutely no surprise to learn that in the geometric algebra of Euclidean three space precisely the same two sided

description of *rotations*[8] is used which made quaternions so precious to Hamilton:

$$m \to Rm\widetilde{R}, \text{ with } R\widetilde{R} = 1, \quad (49)$$

where now

$$R \equiv \pm \exp(-\tfrac{1}{2}b\vartheta), \quad (50)$$

and $b$ is the unit bivector characterizing the plane in which the rotation takes place. I emphasize that Eq. (49) not only applies to vectors, but to *all* elements of (2), because of the property $R\widetilde{R} = 1$. But even more, the *rotors* R commute by construction [compare Eqs. (1-3), (48) and (50)] with the origin and infinity null vectors $\vec{n}, \vec{\bar{n}}$, hence rotations about the origin of conformal points (18) or higher grade elements representing according to section 3.2 circles, lines, spheres, planes and volumes have exactly the same form as the rotation in Eq. (49).

This makes rotations of any geometric entity incredibly easy and also in computer implementations the same rotor R applied to all multivector objects representing geometric entities achieves their rotations, which greatly simplifies the programming.

A very special aspect of conformal geometric algebra is, that it implements *translations* in the very same way as rotations in Eq. (49) which may in some degree be even anticipated considering that now multiplication is related to a measure of distance as in Eq. (20). The translation operator, the *translator T* for translating a multivector by the Euclidean three space vector $\vec{a}$ is

$$T \equiv \exp(\tfrac{1}{2}\vec{n}\vec{a}) = 1 + \tfrac{1}{2}\vec{n}\vec{a}. \quad (51)$$

All higher order terms of the expansion in Eq. (48) of *T* in powers of $\tfrac{1}{2}\vec{n}\vec{a}$ vanish, because of Eq. (6), i.e. the null property of $\vec{n}$.

So for translating e.g. the point $\vec{X}$ we simply write in full analogy to (49)

$$\vec{X} \to T\vec{X}\widetilde{T}, \quad (52)$$

again with

$$T\widetilde{T} = 1. \quad (53)$$

The next step is to combine rotors and translators to achieve rotations about arbitrary centers. For that we first translate from a desired center of rotation $\vec{a}$ back to the origin, using the inverse of *T*, i.e. $\widetilde{T}$, then rotate and finally translate with T back to the center position $\vec{a}$:

$$\vec{X} \to TR\widetilde{T}\vec{X}T\widetilde{R}\widetilde{T} = R'\vec{X}\widetilde{R}', \quad (54)$$

with the rotor $R'$ for rotations about $\vec{a}$

$$R' \equiv TR\widetilde{T} \text{ and } R'\widetilde{R}' = 1, \quad (55)$$

as in Eq. (49).

Finally we can freely combine angle $\vartheta$ rotations $R'$ about any center $\vec{a}$ with translations T to any other position $\vec{b}$ in space to give us combined motion-rotation operators also called *motors* D

$$D \equiv T(\vec{b})R'(\vartheta, \vec{a}), \quad \text{with } D\widetilde{D} = 1. \quad (56)$$

Applying a motor D, e.g. to a conformal point $\vec{X}$ would just again be

$$\vec{X} \to D\vec{X}\widetilde{D}. \quad (57)$$

Because motors D commute by construction with the infinity vector $\vec{n}$ and because of $D\widetilde{D}=1$, the description (57) of combined translations and rotations by motors D also applies to the other geometric entities of section 3.2 (circles, lines, spheres, planes and volumes, etc.)

After knowing about magnitudes, the inverse and ways to move around geometrid objects, the question is for how to combine, (e.g. two lines to form a plane) lower dimensional objects like two lines, or a line and a plane, etc. to form higher dimensional objects, like as planes or the whole space, respectively. In Eqs. (21), (30), (32) we already used the outer product of conformal points to create higher dimensional geometric entities, like pairs of points, lines, spheres, etc. And indeed the outer product of Eq. (41) is together with the contractions of Eqs. (43) and (44) exactly what we need in order to describe the set theoretic union of points or *join* of conformal subspaces, which are represented by single grade multivectors (*blades*) as explained in Eq. (26). Equation (26) applies not only to the three dimensional Euclidean sub-algebra, but in fact to the whole conformal algebra. Let us assume, that we have two conformal subspaces represented by the blades W and V, with the possibility of a common subspace blade factor M:

$$W = W' \wedge M, \text{ and } V = M \wedge V'. \tag{58}$$

Then the join J of these two subspaces will simply be

$$J \equiv W \wedge (M^{-1} \rfloor V). \tag{59}$$

This is without problem, as long as the common blade $M$ does not include the vector factors $\bar{\vec{n}}$ or $\vec{n}$ by themselves. (If it contains both, i.e. $N \equiv \vec{n} \wedge \bar{\vec{n}}$, then we have no problem, because according to Eq. (10), $N$ is inverse to itself.) So for example in the case of an M of the form of Eq. (30)

$$M = \vec{A}_1 \wedge \vec{A}_2 \wedge \vec{n} = M' \wedge \vec{n}, \tag{60}$$

(with $M' \wedge \bar{\vec{n}} \neq 0$) we simply have to replace in Eq. (59)

$$M^{-1} \rfloor V \rightarrow -N(\bar{\vec{n}} \wedge (M'^{-1} \rfloor V)). \tag{61}$$

In the event of

$$M = M' \wedge \bar{\vec{n}}, \text{ and } M' \wedge \vec{n} \neq 0, \tag{62}$$

we have to replace in (59)

$$M^{-1} \rfloor V \rightarrow N(\vec{n} \wedge (M'^{-1} \rfloor V)). \tag{63}$$

Naturally the easiest case is $W$ and $V$ to be disjoint. Then we simply have

$$J \equiv W \wedge V. \tag{64}$$

Examples of the last case are Eqs. (21), (30), (32), etc.

The common subspace blade factor $M$ of (58) is the set theoretic *meet* or *intersection* of the two subspaces $W$ and $V$. Once we know the join of two subspaces, we can in turn calculate the meet as

$$M = (V \rfloor J^{-1}) \rfloor W. \tag{65}$$

In case that $V$ includes one of the vector factors $\bar{\vec{n}}$ or $\vec{n}$, i.e. $V = V'' \wedge \bar{\vec{n}}$ or $V = V''' \wedge \vec{n}$, analogous replacements apply to the product $(V \rfloor J^{-1})$ of Eq. (65), which were made for $M^{-1} \rfloor V$ in Eq. (63) or in Eq. (61), respectively.

Another important set theoretic operation is the projection of a subspace $B$ onto another (single grade) subspace, which we denote here as $A$, represented according to Eq. (26) by corresponding blades $B$ and $A$, respectively. In the case of the whole space we would simply have $A = I$. This is achieved[25] by

$$P_A(B) \equiv (A \rfloor B) \rfloor B. \tag{66}$$

By the bilinearity of the left contraction, which is based on the bilinearity of the multivector product and of the grade selection operations involved, this can be extended to projecting whole multivectors *m* [understood according to Eq. (35) as collections of subspaces] onto *B*:

$$P_A(m) \equiv (m \rfloor B) \rfloor B . \tag{67}$$

To conclude this selective vista of ways to manipulate geometric entities, which are represented by multivectors, let us look closer at the properties of the conformal representation of a line as in Eq. (30) utilizing Eq. (21). Eq. (30) produces

$$V_{line} = \vec{a}_1 \wedge \vec{a}_2 \vec{n} + (\vec{a}_2 - \vec{a}_1) N . \tag{68}$$

$\vec{a}_1 \wedge \vec{a}_2$ is the socalled *moment* bivector of the line and $\vec{a}_2 - \vec{a}_1$ is its vector of diretion. From Eq. (68) we can calculate the distance vector of any point $\vec{x}$ from the line[16]

$$\vec{d} \equiv \left(\vec{x} \wedge (\vec{a}_2 - \vec{a}_1) - (\vec{a}_1 \wedge \vec{a}_2)\right) \frac{1}{(\vec{a}_2 - \vec{a}_1)} . \tag{69}$$

This is a direct way of calculating the distance of any point $\vec{x}$ from the line, only using the two components of $V_{line}$ conformally representing the line.

Now we have learnt enough about five dimensional conformal geometry to get a glimpse of its elegant multivector representations of geometric entities and its algebraically simple but wide ranging computational tools to freely manipulate these geometric entities.

The next section will therefore be completely devoted to the way the new Java package GeometricAlgebra implements the geometric multivector entities as objects (classes) and how it allows to perform the geometric manipulations by means of the methods of these objects.

**4. The New Java Package "GeometricAlgebra"**

This section gives brief information on the Java object classes contained in the package GeometricAlgebra, listing the constructur of each class together with the available methods. Downloading KamiWaAi 0.0.1 beta from the KamiWaAi index page mentioned in the introduction automatically downloads the binary code of the package GeometricAlgebra as well. Simply inspect the sub-folder GeometricAlgebra to find the binary code!

**4.1 Class ComplexNumber**
This is the public class ComplexNumber with the constructor: ComplexNumber(double realPart, double imaginaryPart). It contains the methods listed in Table 5.

Table 5 Methods of class ComplexNumber.

| |
|---|
| double RealPart() |
| double ImaginaryPart() |
| double Magnitude() |
| void setRePart(double rp) |
| void setImPart(double im) |
| ComplexNumber add(ComplexNumber cn) |
| ComplexNumber sub(ComplexNumber cn) |
| ComplexNumber mult(ComplexNumber cn) |

**4.2 Class ComplexQuat**
This is the public class ComplexQuat with the constructor: ComplexQuat(ComplexNumber[] cq). It contains the methods listed in Table 6.

Table 6 Methods of class ComplexQuat.

| |
|---|
| ComplexNumber getScPart() |
| ComplexNumber getScPart() |
| void setScPart(ComplexNumber sp) |
| void setBvPart(ComplexNumber[] bp) |
| ComplexQuat add(ComplexQuat cq2) |
| ComplexQuat sub(ComplexQuat cq2) |
| ComplexQuat mult(ComplexQuat cq2) |

### 4.3 Class MultiVector

This is the public class MultiVector with the constructor: MultiVector(ComplexQuat[] mv). Table 7 shows its methods.

Table 7 Methods of class MultiVector.

|  |
| --- |
| ComplexQuat getnbarPart() |
| ComplexQuat getnhnbPart() |
| ComplexQuat getScPart() |
| ComplexQuat getnPart() |
| MultiVector get3DMVector() |
| MultiVector getGrade(int g) |
| double magnitude() |
| MultiVector normalize() |
| MultiVector reverse() |
| void setScPart(ComplexQuat sp) |
| void setnPart(ComplexQuat np) |
| void setnbarPart(ComplexQuat nbarp) |
| void setnhnbPart(ComplexQuat nhnbp) |
| void setZero() |
| MultiVector add(MultiVector mv2) |
| MultiVector sub(MultiVector mv2) |
| MultiVector mult(MultiVector mv2) |
| MultiVector Powerof(int power) |
| double ScProd(MultiVector mv2) |
| MultiVector OutProd(MultiVector mv2) |
| MultiVector Lcontract(MultiVector mv2) |
| MultiVector Rcontract(MultiVector mv2) |
| MultiVector multSc(double factor) |
| void show() |

### 4.4 Class fMV

This is the public class fMV, a collection of frequently used multivectors. The constructor is fMV(). Its methods are listed in Table 8.

Table 8 Methods of class fMV.

| | |
| --- | --- |
| MultiVector n() | MultiVector I() |
| MultiVector nbar() | MultiVector e1() |
| MultiVector N() | MultiVector e2() |
| MultiVector One() | MultiVector e3() |

### 4.5 Class Line

This is the public class Line with the constructor: Line(MultiVector L2). Its methods are listed in Table 9.

Table 9 Methods of class Line.

|  |
| --- |
| MultiVector Ltrivector() |
| void setColor( Color c ) |
| MultiVector getMoment() |
| MultiVector getLineVector() |
| MultiVector LinePoint(double par) |
| double[] linepoint3D(double par) |
| void drawL(Graphics g) |
| void drawLZ(Graphics g) |
| void setBasePoint(MultiVector basepoint, int p12) |
| void remake(MultiVector movedBP, int bpno) |
| MultiVector PDistance(double[] picked) |
| MultiVector Pdistance2D(double[] point, MultiVector plane) |
| double distanceXY(double[] point) |
| double distanceYZ(double[] point) |

**4.6 Class Circle**

This is the public class Circle with the constructor: Circle(MultiVector L3). Its methods are listed in Table 10.

Table 10 Methods of class Circle.

| |
|---|
| MultiVector Ctrivector() |
| double radius() |
| MultiVector Center() |
| double[] center() |
| MultiVector Cplane() |
| MultiVector generator(int inc) |
| MultiVector PonC() |
| double[][] cpoints(int inc) |
| void setColor( Color c ) |
| void drawC(Graphics g) |
| void drawCZ(Graphics g) |
| void setBasePoint(MultiVector basepoint, int p123) |
| void remake(MultiVector movedBP, int bpno) |

**4.7 Class Sphere**

This is the public class Sphere with the constructor: Sphere(MultiVector L4). It contains the methods listed in Table 11.

Table 11 Methods of class Sphere.

| |
|---|
| double radius() |
| MultiVector Center() |
| double[] center() |
| MultiVector generator(int inc, MultiVector plane) |
| MultiVector PonS() |
| MultiVector PonSsouth() |
| Vector snet(int inc) |
| void setColor( Color c ) |
| void drawC(Graphics g) |
| void drawCZ(Graphics g) |
| void setBasePoint(MultiVector basepoint, int p1to4) |
| void remake(MultiVector movedBP, int bpno) |
| void recenter(MultiVector NewCenter) |
| MultiVector MeetLine(Line line) |
| int MeetNo(Line line) |
| Vector MeetLine3D(Line line) |

**4.8 Class Euc3Vector**

This is the public class Euc3Vector with the constructor: Euc3Vector(double[] vec). Presently it contains only one method: MultiVector conv2MV().

**5. Conclusion**

Finally a few historical remarks are in order. Exactly 170 years ago, the Jablonowski Gesellschaft expressed its interest in an essay by the German mathematician, philosopher and diplomat Gottfried Wilhelm Leibnitz (1646-1716) by offering a prize for the further development of Leibnitz' ideas. Leibnitz had written: "… this new characteristic, which follows the visual figures, cannot fail to give the solution, the construction, and the geometric demonstration all at the same time, and in a natural way and in one analysis, that is, through determined procedure." The one mathematician who competed for the price and promptly won it in 1846 was Herrmann Grassmann, the creator of "a new branch of mathematics", the linear extension theory.[9],[21],[26]

Yet Grassmann's revolutionizing insights suffered "colossal neglect"[9] during his lifetime. Therefore he writes in the preface of his 1862 "Extension Theory": "For I have confidence that the effort I have applied to the science reported upon here, which has occupied a considerable span of my lifetime and demanded the most intense exertions of my powers, is not to be lost … I know, that even if this work as well should lie idle yet another seventeen years or more without influencing the living development of science, a time will come when it will be drawn forth from the dust of oblivion and the ideas laid down here will bear fruit … For truth is eternal, it is divine; and no phase in the development of truth … can pass away without a trace. It remains even if the garments in which feeble men clothe it fall into dust."[26]

In particular section 3 of this paper shows some of the modern continuation of the ideas and creations of Leibnitz and Grassmann. As for the objects used by computers to represent and calculate geometry in software[27] as KamiWaAi, we now finally reach a stage to consistently picture, think, program, calculate and visualize by means of geometric objects. The manipulations of these geometric objects are well "determined procedures" *within* the native algebra of the geometric objects themselves.

**Acknowledgements**

I thank God my creator for the joy of doing research in the wonders of his works: "He (Christ) is before all things, and in him all things hold together."[28]

I thank my wife for continuously encouraging my research. I thank my three year old son Joshua for inventing the Platonic snowman of Fig. 2. I thank the whole Department of Mechanical Engineering at Fukui University for generously providing a suitable research environment for me, especially Prof. Takeshita. I finally thank my friend K. Shinoda, Kyoto for his prayerful personal support.

*L'on voit que Jésus-Christ, achevant ce que Moïse avait commencé, a voulu que la divinité fût l'objet, non seulement de notre crainte et de notre veneration, mais encore de notre amour et de notre tendresse.* G.W. Leibnitz[29]